%% file: main.tex
\documentclass{ws-procs11x85}
\usepackage{ws-procs-thm}           

\usepackage{xcolor}
\usepackage{xurl}
\usepackage{booktabs}
\usepackage{array}

\newcommand{\eg}{\emph{e.g.,}\;}

\newcommand\toolname{\texttt{ETHOS}\;}

\begin{document}

\title{\toolname: Towards a Modular Ethics Framework for Clinical Multi-Agent Systems}

\author{Rakesh Sharma$^{1,\dag}$, Sydney Pugh$^{2,\dag}$, Cameron Beeche$^{1}$,
Pankhuri Singhal$^{3}$, Rachel Wu$^{2}$, Margaret Eby$^{4}$, Jeffrey Duda$^{1}$,
James Gee$^{1}$,  Kyra O'Brien$^{5}$,
Hersh Sagreiya$^{1}$, Marina Serper$^{6}$, Victoria Gershuni$^{7}$, Angela Bradbury$^{8,4}$, Anurag Verma$^{3}$,
Eric Eaton$^{9}$, Kevin B. Johnson$^{2,9}$ and Walter Witschey$^{1}$}
\address{$^{1}$Department of Radiology, University of Pennsylvania,
Philadelphia, PA 19104, USA\\[3pt]
$^{2}$Department of Biostatistics, Epidemiology, and Informatics,
University of Pennsylvania, Philadelphia, PA 19104, USA\\[3pt]
$^{3}$Department of Medicine, Division of Translational Medicine and Human Genetics, University of Pennsylvania, Philadelphia, PA 19104, USA\\[3pt]
$^{4}$Department of Medical Ethics and Health Policy,
University of Pennsylvania, Philadelphia, PA 19104, USA\\[3pt]
$^{5}$Department of Neurology, University of Pennsylvania, Philadelphia, PA 19104, USA\\[3pt]
$^{6}$Division of Gastroenterology and Hepatology, University of Pennsylvania,
Philadelphia, PA 19104, USA\\[3pt]
$^{7}$Department of Surgery, University of Pennsylvania, Philadelphia, PA 19104, USA\\[3pt]
$^{8}$Division of Hematology-Oncology, Department of Medicine,
University of Pennsylvania, Philadelphia, PA, USA\\[3pt]
$^{9}$Department of Computer and Information Science, School of Engineering and Applied Science,
University of Pennsylvania, Philadelphia, PA, USA\\[3pt]
$^{\dag}$These authors contributed equally.\\
Corresponding Author E-mail: rakesh.sharma@pennmedicine.upenn.edu}
\begin{abstract}

The rapid adoption of large language models has enabled the development of clinical multi-agent systems (MAS) capable of integrating multimodal patient data and supporting increasingly complex clinical decision-making. However, the deployment of these systems in real-world healthcare settings raises critical ethical concerns related to safety, fairness, accountability, transparency, and patient trust. While numerous organizations, including the World Health Organization, the National Academy of Medicine, and the FUTURE-AI consortium, have proposed ethical frameworks and governance principles for healthcare AI, these efforts remain largely conceptual. A major gap persists between ethical guidance and its operational implementation within deployed AI systems.

To address this challenge, we present ETHOS (Ethics and Trust through Hierarchical Oversight System), a modular ethics framework designed as a governance meta-agent that can be integrated with any existing multi-agent system without requiring changes to its underlying architecture. ETHOS translates stakeholder-informed ethical requirements into executable runtime oversight through a layered governance approach consisting of deterministic checks, contextual reviews, and a final ethics critic. These components continuously evaluate intermediate reasoning steps and final outputs, enabling the system to identify ethical risks, request revisions, or suppress responses that fail predefined safety and trustworthiness criteria.

We demonstrate ETHOS within a hepatology clinical decision-support MAS that combines electronic health record data, medical imaging, predictive models, and clinical practice guidelines. Results show that ETHOS improves decision reliability by detecting incomplete, inconsistent, or out-of-scope evidence and appropriately increasing abstention when safe recommendations cannot be supported. By embedding ethical governance directly into system operation, ETHOS provides a practical and auditable mechanism for transforming high-level AI ethics principles into deployable safeguards. More broadly, ETHOS offers a generalizable approach for enabling trustworthy deployment of clinical and other high-stakes multi-agent AI systems.
\end{abstract}


\textit{Preprint of an article submitted for consideration in Pacific Symposium
on Biocomputing \textcopyright\ 2027 World Scientific Publishing Company. \url{https://psb.stanford.edu/}}

\copyrightinfo{\copyright\ 2024 The Authors. Open Access chapter published by World Scientific Publishing Company and distributed under the terms of the Creative Commons Attribution Non-Commercial (CC BY-NC) 4.0 License.}

\newpage
\input{intro}
\input{related}
\input{stakeholder}
\input{methods}
\input{liver}
\input{results}
\input{discussion}

\section*{Acknowledgments}

M.E. acknowledges support from the National Human Genome Research Institute
training grant T32HG009496. C.B. acknowledges support from the National Heart,
Lung, and Blood Institute under Award Number F31HL182332. S.P. and K.B.J. acknowledge
support from NIH DP1-LM-014558. W.W. acknowledges support from NIH
P41-EB029460, R01-HL169378, R01-HL171709, UL1-TR001878, OT2-OD038048, and
R21-EB036734. E.E. acknowledges that this material is based upon work supported by the U.S. National Science Foundation under Cooperative Agreement No. 2433450. The content is solely the responsibility of the authors and does
not necessarily represent the official views of the National Institutes of
Health or the National Science Foundation.

\section*{Supplementary Material}
All appendices can be found {\scriptsize\url{https://github.com/PennMultimodalAI/ETHOS-Supplementary/blob/main/sup_materials.pdf}}.

\bibliographystyle{ws-procs11x85}
\newpage
\bibliography{references}


\end{document}

%% file: intro.tex
\section{Introduction}
\label{sec:intro}

Clinical decision making inherently requires the holistic integration of diverse, multimodal information across a patient's course of care. Recently, clinical AI development has increasingly leveraged foundation models to navigate this complexity by creating clinical \emph{multi-agent} systems (MAS)~\cite{ferber2025development, ferber2026towards, lievin2026towards}. These systems decompose a clinical query into discrete steps carried out by autonomous agents with access to specialized tools (\eg image segmentation or risk prediction models).
In practice, however, the shortcomings of these systems and their underlying models become apparent: 
(1) they exhibit unpredictable and potentially unsafe behavior when encountering novel clinical scenarios or complex queries;
(2) they can display answers and behaviors that belie their ethical application to medicine; and (3) they lack established best practices, ethics frameworks, and co-design processes for the real-world ethical challenges of using AI in clinical settings. \cite{xie2026ethical, marco2024multinational, jain2024artificial, lee2025vulnerability} 

These shortcomings highlight a critical gap in clinical AI governance, motivating a need for \emph{integrated ethics enforcement}: ethical requirements imposed on the MAS as it runs.
Several frameworks supply conceptual guidance for this, including the World Health Organization's guidance on the ethics and governance of AI for health~\cite{guidance2021ethics, world2024ethics}, FUTURE-AI~\cite{lekadir2025future}, and codes of conduct from the National Academy of Medicine~\cite{adams2025advancing} and the American Medical Association~\cite{american1848code}.
These frameworks provide categorical constructs (\eg inclusivity and equity, autonomy) rather than specifications, however, and the concrete concerns falling under them depend on the clinical application and the setting in which it is deployed. 
Identifying the \emph{specific} ethical concerns requires engaging the clinicians, informaticians, privacy staff, and administrators who work in that setting, and revisiting them as practice changes. 

Translating these stakeholder-defined concerns into runtime checks places three demands on the system. First, checks must run during inference, applied to the intermediate results produced at each step as well as to the final assembled output. Second, concerns must be operationalized as concrete checks attached to the individual agents and tools whose behavior they constrain, since a construct such as equity does not by itself specify what to verify, or where. Third, the findings must be consequential -- capable of sending a draft answer back for revision, or of preventing it from reaching the clinician at all.

We present \toolname (\textbf{E}thics and \textbf{T}rust through \textbf{H}ierarchical \textbf{O}versight \textbf{S}ystem), a modular ethics meta-agent framework for clinical MAS. 
\toolname implements core ethics checks that are reused across the agents and tools of a MAS, with what they examine and how they are parameterized supplied by a per-application configuration that serves as an executable governance policy.
The design of these checks is informed by ethical concerns raised by clinical stakeholders through an iterative co-design process (Section~\ref{sec:stakeholder}). 
Concerns are operationalized at two tiers. Within each sub-agent, \emph{pre-specified} checks apply a fixed, predetermined criterion on every invocation of a tool, while \emph{contextual} checks require case-specific reasoning over the individual patient's clinical context; their findings are appended to the sub-agent's response and re-evaluated as evidence is compiled across sub-agents. At the orchestrator level, an \emph{ethics critic} agent reviews the drafted answer for potential patient harm and returns feedback summarizing any outstanding ethical issues, which the orchestrator must resolve -- by revising its answer, or suppressing it -- before that answer reaches the clinician.

To evaluate \toolname, we apply the framework to 
a hepatology MAS that combines multimodal patient data (\eg laboratory results from the EHR and CT images) with access to hepatology clinical practice guidelines. The agents collaboratively screen for liver-related diseases, assess disease severity and risk factors, and propose evidence-based treatment and management strategies. 

In summary, this paper makes the following contributions:
\begin{itemize}
    \item \toolname, a modular ethics meta-agent framework for clinical MAS that enforces stakeholder-informed ethical checks at runtime.
    \item A preliminary mapping of stakeholder-elicited ethical concerns to established ethics frameworks and to the concrete checks implemented in \toolname. 
    \item An application of \toolname to a guideline-informed hepatology diagnosis and treatment support MAS.
    \item An evaluation of \toolname quantifying the effect of integrated ethics enforcement on the diagnostic accuracy of the hepatology MAS and on the share of decisions the system answers, stratified by whether the evidence available to the system was complete.
\end{itemize}


%% file: related.tex
\section{Related Work}
\label{sec:related}








Existing ethical guidance for clinical and agentic AI is largely declarative rather than operational. The WHO framework introduced above supplies constructs such as accountability and transparency but does not specify how a deployed system should be checked against them~\cite{guidance2021ethics,world2024ethics}. Other guidance efforts extend this declarative pattern to medicine specifically, including FUTURE-AI's consensus checklist for trustworthy AI in healthcare~\cite{lekadir2025future} and the National Academy of Medicine's code of conduct for AI in health~\cite{adams2025advancing}; like the WHO framework, these supply high-level constructs rather than a mechanism for checking a deployed system against them. Preliminary stakeholder-facing work confirms this: interviews with 23 healthcare professionals on AI-driven clinical decision support for resource allocation surfaced concerns spanning equity, transparency, shifting clinical accountability, and data governance which varied by professional role and care setting rather than mapping cleanly onto a fixed set of principles~\cite{Elgin2024}. A systematic review of the broader debate on AI-supported clinical ethical decision-making similarly found that justice, explicability, and human-AI interaction remain comparatively underexplored relative to autonomy, with little consensus on how any of these principles should be implemented~\cite{Benzinger2023}. A recent benchmark found that direct application of LLM-generated clinical recommendations risked severe patient harm in up to 24.6\% of cases, driven mostly by errors of omission, and that multi-agent approaches reduced but did not eliminate this risk~\cite{wu2026firstnoharmmedicalsafety}. A parallel conceptual literature has begun mapping classical AI ethics principles onto agentic systems specifically, arguing that the broader agency, greater impact on human autonomy, and deeper human-technology entanglement of AI agents amplify existing concerns around transparency, fairness, non-maleficence, accountability, and privacy rather than introducing categorically new ones~\cite{faucris.360624675}. Like the WHO guidance, this analysis identifies which principles agentic systems strain without proposing a mechanism for enforcing them at runtime.

A smaller technical literature addresses enforcement directly, though largely outside clinical multi-agent settings and framed around safety and access control rather than the broader stakeholder-elicited concerns we target. GuardAgent checks a target agent's inputs and outputs against user-specified guard requests by generating and executing guardrail code, evaluated in part on a healthcare access-control benchmark~\cite{xiang2025guardagentsafeguardllmagents}, and AgentSpec compiles natural-language rules into runtime constraints that can intervene on an agent's actions before execution completes~\cite{wang2025agentspeccustomizableruntimeenforcement}. These systems establish that runtime enforcement of externally specified rules is feasible for LLM agents, but neither incorporates the multimodal, multi-step evidence aggregation characteristic of clinical MAS, and their rule sets are fixed safety policies rather than concerns elicited from clinical stakeholders and revisited as practice changes.

Taken together, the volume of ethical guidance for clinical and agentic AI stands in sharp contrast to the amount of work translating that guidance into runtime, multimodal enforcement; \toolname is designed to close that gap.

%% file: stakeholder.tex
\section{Stakeholder Engagement}
\label{sec:stakeholder}

We utilized an iterative mixed-methods design to obtain the input of multiple stakeholders in clinical care settings to collect input on the impact, benefit and challenges to application of ethics in diverse real-world clinical settings throughout the design and evaluation of a MAS. The stakeholders consisted of ten experts across the healthcare ecosystem, including providers as well as representatives from clinical operations and leadership, clinical informatics, and privacy, legal, and governance. Stakeholders convened at regular, three month intervals for structured meetings where evolving use cases and feedback on the design process are presented by relevant members of the \toolname engineering team. Transcripts from these meetings were then coded using rapid qualitative assessment (RQA), which uses structured summaries of interviews or focus groups organized into templated matrices aligned with the study’s key domains. A team-based analytic process allowed for rapid comparison across cases, iterative refinement of themes, and early identification of patterns relevant to implementation or policy decisions~\cite{vindrola2020rapid}. RQA prioritizes timely insight over exhaustive coding, making it particularly well suited for grant-funded, applied, or implementation-focused research~\cite{hamilton2013rapid}.

The results of this RQA were then mapped onto the ethics frameworks per: FUTURE-AI, the National Academy of Medicine’s Health Care Artificial Intelligence Code of Conduct, the American Medical Association’s framework for Trustworthy Augmented Intelligence in Health Care, and the World Health Organization's guidance on the ethics and governance of AI. Feedback is linked to principles across these frameworks both by frequency (how often each concern was mentioned across the stakeholder meetings) and qualitatively (by linking substantive comments to principles). This mapping process allows for researchers across ethics and engineering to link feedback to specific ethics principles and subsequently organize them according to design and development timelines, ensuring relevant feedback is incorporated at the appropriate points in the model lifecourse. 

Table~\ref{tab:ethos-mapping} presents this mapping alongside the \toolname components each concern motivated. 
The mapping is neither one-to-one nor confined to a single point in the system, and this shaped the design of \toolname. Some are verifiable against a criterion fixed in advance of any patient case: quality control of tool inputs and outputs yielded checks for temporal discordance between a scan and its associated laboratory values, and for whether absolute attenuation thresholds are applicable to a contrast-enhanced study. Others cannot be evaluated without reasoning about the individual patient, like when demonstrating local clinical validity requires judging whether the population a reference threshold was derived from is appropriate for the case at hand. Still others are properties of a complete answer rather than of any single tool invocation; communicating the role of AI in a clinical decision can only be assessed once an answer has been drafted, and is therefore addressed by the ethics critic alone. Local clinical validity, by contrast, motivated components at all three levels. These differences in what verification requires, and in where it must occur, motivate the tiered structure of \toolname described in Section~\ref{sec:methods}.


\begin{table}[t]
\tbl{Mapping of stakeholder requirements to FUTURE-AI principles, illustrative ETHOS review excerpts, and triggered ETHICS checks.\label{tab:ethos-mapping}}
{\scriptsize
\setlength{\tabcolsep}{12pt}
\renewcommand{\arraystretch}{1.5}
\begin{tabular}{@{}p{1.15in}p{0.60in}p{1.95in}p{0.75in}@{}}
\toprule
\textbf{Ethical concerns} & \textbf{FUTURE-AI mapping} & \textbf{ETHOS review} & \textbf{ETHICS check triggered} \\
\midrule
 Evaluate and demonstrate local clinical validity. 
&  Universality
& \emph{``\ldots the primary HU threshold was derived from cohorts not confirmed to be representative of Black/African American patients; this limits confidence in the finding for this specific patient and independently supports an indeterminate rather than affirmative classification\ldots''} \newline\newline \emph{``\ldots FIB-4 $=$ 0.25 is below the 1.3 low-risk threshold, but guidelines explicitly flag suboptimal FIB-4 performance in patients $<$ 35 years (this patient is 32), reducing reliability of the low score\ldots''}
&  \textbf{Pre-specified}: Reference relevance; \textbf{Contextual}; \textbf{Ethics Critic} \tabularnewline
\addlinespace
 Define mechanisms for quality control of AI inputs and outputs.
&  Traceability
& \emph{``\ldots Critical caveats: all FIB-4 labs were collected 24 to 37 days post-scan; no elastography available; albumin/bilirubin absent. Fibrosis cannot be confirmed or excluded without second-tier NIT\ldots''} \newline\newline \emph{``\ldots absolute HU thresholds are inapplicable on contrast CT\ldots''}
&  \textbf{Pre-specified}: Temporal discordance; CT contrast reliability; \textbf{Contextual} \tabularnewline
\addlinespace
 Rules regarding how to communicate use of AI in medical decision making to patients.
&  Fairness
& \emph{``\ldots This assessment is provided to assist the treating physician; final diagnostic and management decisions rest with the responsible clinician\ldots''} \newline\newline \emph{``\ldots Physician review required before clinical action\ldots''}
&  \textbf{Ethics Critic} \tabularnewline
\bottomrule
\end{tabular}}
\end{table}

%% file: methods.tex
\section{Methods}
\label{sec:methods}

\subsection{\toolname}

We developed ETHOS, an external governance meta-agent that provides ethical and reliability oversight for clinical multi-agent systems (MASs) without requiring modification of the underlying architecture. ETHOS attaches to a host MAS through a lightweight connector that interfaces with the system's tool registry, agent finalization stage, and response revision pathway, allowing the planner, specialist agents, and tools to remain unchanged. Governance is implemented through a layered architecture comprising pre-specified, contextual, and critic components. The pre-specified layer interposes deterministic safety and reliability checks at tool boundaries and executes them automatically whenever associated tools are invoked. Each check generates a structured finding containing the detected issue which is appended to the tool output to the sub-agent LLM. 

To complement deterministic validation, ETHOS incorporates an adaptive governance process consisting of contextual review, and ethics adjudication. Before a sub-agent finalizes its output, the contextual layer prompts the agent to assess whether the specific clinical scenario warrants additional scrutiny beyond the predefined rule set and, when appropriate, to perform supplemental analyses or revise its rationale. After all agent outputs are aggregated, another review layer evaluates the compiled response for inconsistencies, omissions, or risks that emerge only through cross-agent interaction. The final stage is an ethics critic that evaluates the complete response using accumulated governance findings together with case-relevant ethical and clinical guidance. The critic assesses the response against a fixed rubric consisting of ethical principles (beneficence and non-maleficence), predefined guardrails, and a safety-trust-patient impact framework. Approval is enforced programmatically and orchestrator is allowed to release its response only when all policy criteria are satisfied. When the response fails review, ETHOS returns the critic to the host MAS for review and re-adjudicates for a bounded number of iterations (default=5); persistent failures result in suppression of the response and generation of an explanatory notice.

\begin{figure}[tb]
    \centering
    \includegraphics[width=1.0\linewidth]{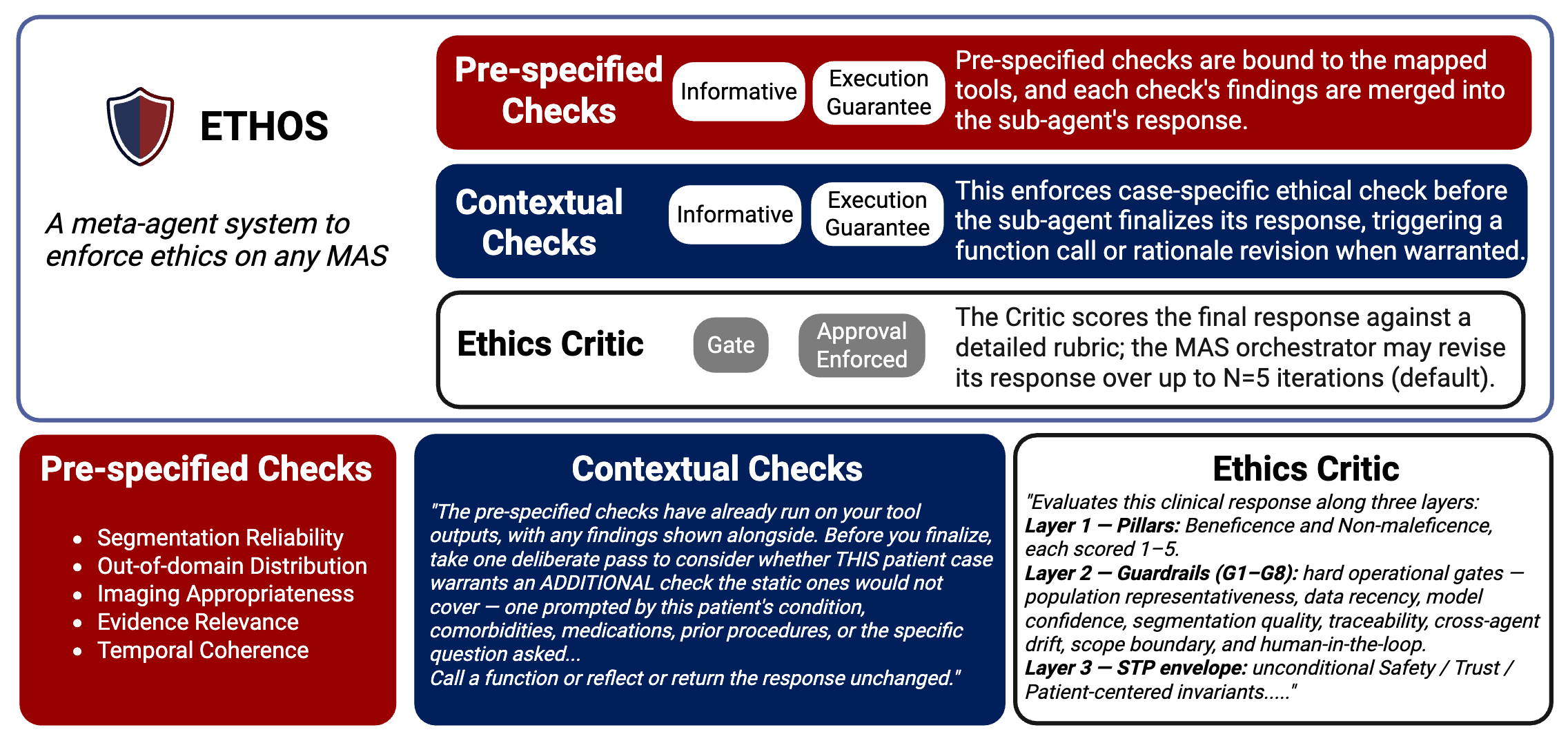}
    \caption{\toolname framework overview.}
    \label{fig:ethos_framework}
\end{figure}

\subsection{Pre-Specified Checks}

\subsubsection{Segmentation Reliability}
ETHOS implements a set of deterministic imaging checks that operate at the boundaries of imaging-analysis tools, with the objective of identifying measurements or predictions derived from potentially unreliable segmentation masks. Three checks target image and measurement integrity:
i) A CT contrast-phase reliability check evaluates the estimated post-contrast acquisition time produced by the contrast-phase classifier and indicates uncertainty in phase assignment. ii) A segmentation-boundary check quantifies the number of voxels contiguous with the boundary of the imaged field of view. Masks that contact the image boundary indicate truncation or incomplete organ capture, rendering derived volumetric, morphologic, and attenuation-based measurements unreliable. 
iii) An abnormal-attenuation check identifies voxels with unusually high Hounsfield unit values within each segmentation mask, flagging the presence of metallic implants, devices, foreign bodies, or focal contrast accumulation that may distort attenuation-derived biomarkers and quantitative features.

\subsubsection{Out-of-Distribution}

Predictive model-based tools typically under-perform when applied to patient cases unlike those which they were developed on \cite{finlayson2021clinician, cohen2021problems}. \toolname therefore provides an out-of-distribution (OOD) check that scores a case by its distance from a reference cohort of the cases a tool was developed on and flags it when that distance exceeds a threshold. The check is model-agnostic, operating on any fixed-length vector representation of a model's inputs or outputs (\eg learned embeddings, hand-crafted features, logits). We support a parametric Mahalanobis distance~\cite{lee2018mahalanobis} and a nonparametric $k$-nearest-neighbor distance~\cite{sun2022knn}, the latter making no assumption about the form of the reference distribution. Thresholds are derived from the reference cohort itself as percentiles of its own score distribution, so selecting the $p$-th percentile calibrates the check to flag approximately $100 - p$ percent of in-distribution cases, making the expected false-positive rate an explicit configuration choice rather than an emergent property of the score. The distance function, reference cohort, and threshold percentile are supplied by configuration; full details are provided in the supplementary material.

\subsubsection{Imaging Appropriateness}
ETHOS additionally validates the applicability of the CT foundation model through three complementary checks that are evaluated jointly. i) An out-of-distribution check uses the OOD detector described above to determine whether an input scan lies outside the training distribution on which model performance was established. 
ii) A region-coverage check verifies concordance between the anatomical region predicted from the scan and the region-specific classifier head used to generate disease predictions. 
ii) A contrast-determinability check assesses whether the model's internal contrast classifier can confidently determine the acquisition phase of the examination. Findings generated by any of these checks indicate that foundation-model outputs may not be trustworthy for the study under evaluation. 

\subsubsection{Evidence Relevance}
When the host MAS incorporates external evidence through retrieved biomedical literature or clinical-practice guidelines, ETHOS applies an evidence-relevance check to assess whether the retrieved material substantively supports the claim for which it is cited. The evaluator determines whether the evidence is directly supportive, tangentially related, or non-supportive of the claim. This check is designed to mitigate illusory grounding.

\subsubsection{Temporal Coherence}

ETHOS evaluates the temporal relationship between laboratory studies and imaging examinations using a deterministic temporal-coherence check. This check is primarily intended for laboratory-derived clinical indices that integrate imaging findings with biochemical measurements, such as the FIB-4 fibrosis score, which combines imaging context with aspartate aminotransferase (AST), alanine aminotransferase (ALT), and platelet values. ETHOS reports these temporal offsets as informational findings while leaving their clinical interpretation to downstream reasoning components.


\subsection{Contextual Checks}
Because a fixed library of deterministic checks cannot exhaustively cover all clinically relevant ethical and reliability considerations, ETHOS incorporates a contextual review layer that performs case-specific governance at two levels. First, immediately before a specialist sub-agent finalizes its output, ETHOS interrupts the agent and returns it to its reasoning loop with a directed prompt asking whether the patient's clinical context requires additional scrutiny beyond the pre-specified checks. When such a consideration is identified, the agent is instructed to address it by either acquiring additional evidence through tool invocation or revising its reasoning and conclusions accordingly. The resulting correction or caveat is incorporated into the final agent output. Second, after sub-agent outputs have been aggregated into a single response, ETHOS performs a cross-agent contextual review prior to ethics adjudication. This review examines the assembled evidence for issues that may emerge only during synthesis, including conclusions that are insufficiently supported by the combined evidence, unresolved inconsistencies between information sources, omission of previously identified caveats, or case-specific considerations apparent only at the aggregate level. The aggregation agent is prompted to re-evaluate the compiled response and revise it when warranted, while leaving it unchanged otherwise. Together, these two mechanisms provide adaptive oversight at both the individual-agent and multi-agent synthesis levels.

\subsection{Ethics Critic}
The final layer of ETHOS is an ethics critic that adjudicates the compiled multi-agent response before release. The critic is implemented as a language-model reviewer that receives three inputs: (i) the draft response generated by the host MAS, (ii) the complete set of pre-specified and contextual findings accumulated during execution, and (iii) case-relevant passages retrieved through dense retrieval from a vector store containing clinical and ethics guidelines. Using these inputs, the critic evaluates the response against a structured ethical rubric and returns a machine-readable assessment. The rubric comprises three components: two ethical pillars, beneficence and non-maleficence, each represented by an ordinal score; a predefined set of explicit guardrails representing prohibited failure modes; and a safety-trust-patient framework that captures broader concerns affecting clinical safety, user trust, and patient welfare. For each assessment category, the critic records supporting rationale, associated governance findings, relevant guideline citations, and actionable recommendations for revision. In contrast to the preceding governance layers, which identify and address individual concerns, the ethics critic performs a holistic evaluation of whether the response as a whole satisfies the ethical and safety requirements for clinical deployment.

The critic's output serves as the basis for a programmatically enforced release decision. A response is approved only when both ethical pillars meet predefined minimum criteria (default 3 of 5), no guardrail violations are identified, and no safety-trust-patient concerns remain unresolved. When approval is withheld, ETHOS returns the response and associated feedback to the host MAS for revision and subsequently re-evaluates the revised draft. This critique-revision cycle continues for a bounded number of iterations or until the response satisfies all approval criteria. If approval is not achieved within the allotted rounds, the response is blocked and withheld from the clinician, and ETHOS instead returns a notice describing the unresolved concerns. The ethics critic therefore acts as the framework's final release gate, transforming governance findings and guideline evidence into an auditable, policy-enforced determination of whether a response can be safely delivered.

%% file: liver.tex
\section{Hepatology Multi-Agent System}
\label{sec:liver_agentic}

\subsection{Agentic system}
The Hepatology Multi-Agent System (MAS) was developed as a clinical decision-support framework for evaluating liver disease in outpatient and specialty hepatology settings where CT imaging and routine laboratory data are available. The system is organized around a central orchestrator that decomposes a clinician's natural-language query into a sequence of subtasks, routes them to specialist agents, and synthesizes the resulting evidence into a unified response. Clinical data acquisition is performed by an \textbf{EHR agent}, which retrieves patient demographics, age at imaging, body-mass index, CT acquisition metadata, and laboratory values required for liver disease assessment, including aspartate aminotransferase (AST), alanine aminotransferase (ALT), and platelet count. Hepatology-specific reasoning is performed by a \textbf{Liver agent}, which calculates the Fibrosis-4 (FIB-4) index\cite{vallet2007fib} and retrieves relevant recommendations from clinical-practice guidelines using semantic search. Together, these agents provide structured clinical and laboratory context for the assessment of hepatic fibrosis, cirrhosis, and steatosis.

Imaging assessment is performed by two complementary specialist agents. The \textbf{RadX agent} analyzes CT examinations by determining contrast phase, segmenting abdominal organs using TotalSegmentator\cite{d2024totalsegmentator, wasserthal2023totalsegmentator}, and extracting quantitative radiomic features with PyRadiomics\cite{van2017computational}. The \textbf{Percival agent} applies a three-dimensional CT foundation model\cite{beeche2025generalizable} to generate imaging-based risk estimates from volumetric CT data. Each specialist operates within a reason-and-act loop and returns structured findings that are aggregated by the orchestrator into an evidence-grounded clinical response. The intended use case of the MAS is to support hepatology evaluation and opportunistic liver disease screening by integrating laboratory, imaging, and guideline-based evidence, while providing a realistic clinical testbed for evaluating the impact of ETHOS governance on multi-agent reasoning in patient-care workflows.


\subsection{Dataset}
All data were obtained from the Penn Medicine Biobank (PMBB)\cite{verma2022penn}, a consented research repository containing longitudinal clinical data, laboratory measurements, imaging studies, and linked electronic health records from patients receiving care within the Penn Medicine health system. All data access was conducted under Institutional Review Board protocol 813913. From this resource, we assembled a cohort of 225,896 CT studies from 72,523 unique patients for development of the Percival CT foundation model classifier heads. For each study, a 768-dimensional image embedding generated by the foundation model served as the input representation. Condition-specific labels were derived from linked electronic health records by assigning a positive label when the corresponding ICD-10 diagnosis code was present within ±30 days of the CT acquisition date. Data were partitioned at the individual level into training (176,313 studies) and validation (49,583 studies) cohorts, and the resulting split was verified to be person-disjoint. Studies were further assigned to anatomical regions for region-specific modeling. The final dataset comprised 116,536 chest studies and 109,070 abdomen-pelvis studies. Region-specific logistic linear-probe classifiers were trained on the training partition.

Evaluation of the ETHOS framework was performed using an independent cohort of 50 patient cases sampled from the held-out validation partition. For each case, the Hepatology MAS was provided access to the patient's CT examination, relevant laboratory measurements, demographics, and imaging metadata. The evaluation task was to diagnose three hepatology-related ICD-10 codes (K74.0 - hepatic fibrosis, K74.6 - liver cirrhosis, and K76.0 - fatty liver) from a single imaging time point, using the corresponding CT study together with the nearest available demographic and laboratory values relative to that imaging date.  





\begin{figure}[tb]
    \centering
    \includegraphics[width=1\linewidth]{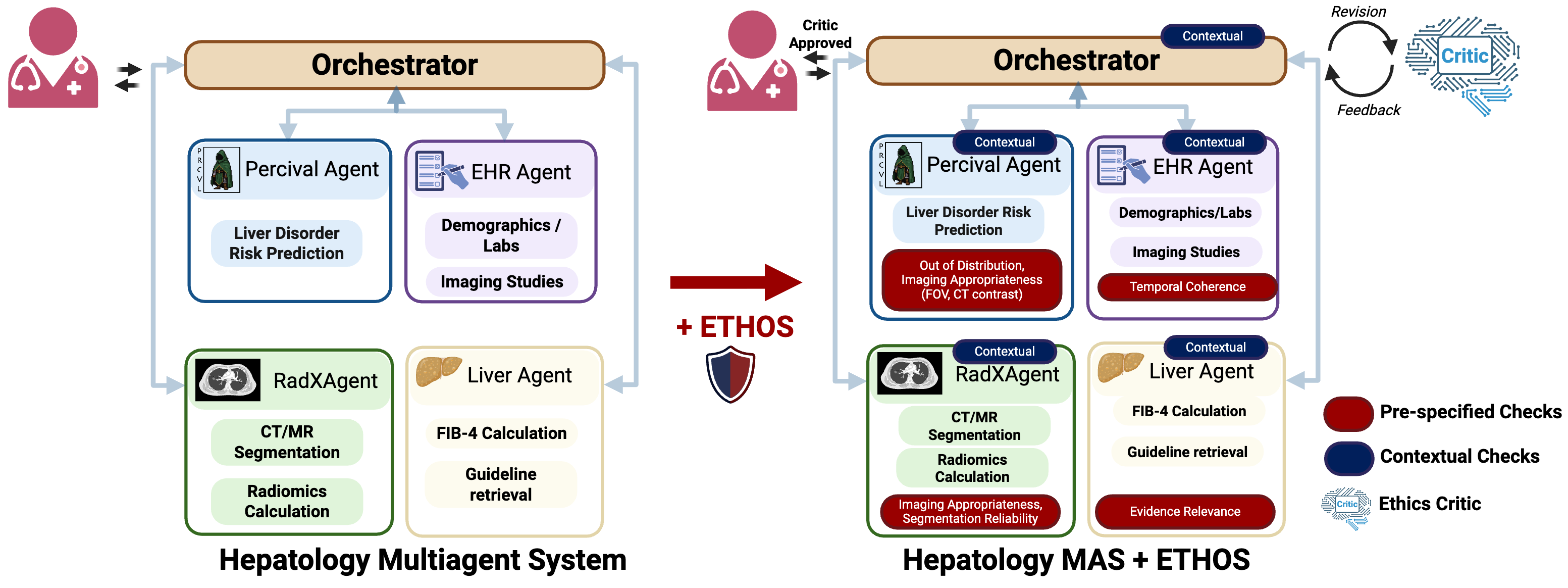}
    \caption{\toolname applied to the hepatology MAS}
    \label{fig:liver_with_ethos}
\end{figure}

%% file: results.tex
\section{Results}
\label{sec:results}

\subsection{Performance with Complete Multimodal Evidence}
This experiment was designed to assess whether \toolname modifies the behavior of the base MAS when all required imaging and laboratory evidence is available and correctly matched, thereby isolating the effect of the critique framework under ideal input conditions. On the 15 decisions with complete, correctly matched imaging and laboratory inputs (5 patients $\times$ 3 ICD codes; Table~\ref{tab:complete-stratum}), \toolname left the abstention rate of the base MAS unchanged for every code: 0\% for K74.0 and K74.6, and 60\% for K76.0. The single decision \toolname altered was a K76.0 case, flipping from a negative to a positive response, matching the ground-truth diagnosis; the Contextual and Ethics Critic checks were responsible for this flip. This indicates that, when the evidence available to the system is complete and consistent with the input requirements of each sub-agent, \toolname's checks rarely modify MAS's original response, and the one intervention observed in this stratum corrected rather than introduced an error.

\begin{table}[tb]
\tbl{Complete-input stratum: all required modalities present and correct, $n=15$ decisions.\label{tab:complete-stratum}}
{\fontsize{9}{10}\selectfont
\setlength{\tabcolsep}{7pt}
\renewcommand{\arraystretch}{1.2}
\begin{tabular}{@{}l c c c c c >{\raggedright\arraybackslash}p{0.82in} >{\raggedright\arraybackslash}p{0.72in} >{\raggedright\arraybackslash}p{1.0in}@{}}
\toprule
Code & MAS Abst. & $+$ETHOS Abst. & $\Delta$ Abst.\ (pp) & Flip \% & Flips (dir.) & Flip$\to$conf.\ correct & Top changing checks \\
\midrule
K74.0 & 0\% & 0\% & 0 & 0\% & --- & --- & --- \\
\addlinespace
K74.6 & 0\% & 0\% & 0 & 0\% & --- & --- & --- \\
\addlinespace
K76.0 & 60\% & 60\% & 0 & 20\% & no$\to$yes: 1 & 1/1 correct (GT pos) & Contextual (1), Ethics Critic (1) \\
\bottomrule
\end{tabular}}
\end{table}

\subsection{Performance under Degraded Multimodal Evidence}
On the 45 decisions in the inadequate-input stratum (15 patients $\times$ 3 ICD codes; Table~\ref{tab:inadequate-stratum}), \toolname raised the abstention rate for every code: from 40\% to 67\% for K74.0, from 47\% to 67\% for K76.0, and from 33\% to 53\% for K74.6. Aggregated across the stratum, abstentions rose from 18 of 45 decisions (40\%) under the base MAS to 28 of 45 (62\%) under \toolname. Most interventions moved a decision toward abstention rather than away from it: all 4 flips on K74.0 and all 3 flips on K76.0 converted a confident response into an indeterminate one, driven primarily by the Imaging Appropriateness and Contextual checks. K74.6 was the exception, where 2 of its 7 flips converted an indeterminate response into a confident negative, both consistent with the ground truth, while the remaining 5 again moved toward abstention. Across the full stratum, only 2 of the 45 decisions were newly resolved to a confident, correct outcome by \toolname, and both occurred on K74.6; the framework's dominant effect under degraded input was to withhold a confident answer rather than to correct one.

\begin{table}[tb]
\tbl{Inadequate-input stratum: a required modality is missing or incorrect, $n=45$ decisions. Abstention is the per-code rate; $\Delta$ Abst.\ is $+$ETHOS $-$ MAS (percentage points).\label{tab:inadequate-stratum}}
{\fontsize{9}{10}\selectfont
\setlength{\tabcolsep}{8pt}
\renewcommand{\arraystretch}{1.2}
\begin{tabular}{@{}l c c c c >{\raggedright\arraybackslash}p{0.7in} >{\raggedright\arraybackslash}p{0.7in} >{\raggedright\arraybackslash}p{1.0in}@{}}
\toprule
Code & \shortstack{MAS\\Abst.} & \shortstack{$+$ETHOS\\Abst.} & $\Delta$ Abst.\ (pp) & Flip \% & Flips (dir.) & Flip$\to$conf.\ correct & Top changing checks \\
\midrule
K74.0 & 40\% & 67\% & $+27$ & 27\% & no$\to$ind: 3 \newline yes$\to$ind: 1 & --- & Imaging Appropriateness (4), Contextual (3) \\
\addlinespace
K74.6 & 33\% & 53\% & $+20$ & 47\% & ind$\to$no: 2 \newline no$\to$ind: 4 \newline yes$\to$ind: 1 & 2/2 correct (GT neg) & Contextual (6), Imaging Appropriateness (5) \\
\addlinespace
K76.0 & 47\% & 67\% & $+20$ & 20\% & no$\to$ind: 3 & --- & Imaging Appropriateness (4), Contextual (2) \\
\bottomrule
\end{tabular}}
\end{table}

\subsection{Diagnostic Performance}
Table~\ref{tab:mas-vs-ethos} reports performance metrics across the full cohort ($N=50$ patients); accuracy, sensitivity, specificity, and F1 are computed only over the cases each system did not abstain on. \toolname increased abstentions for every code, raising the overall abstentions from 40 to 59 decisions (26.7\% to 39.3\% of the cohort), with the largest increase on K76.0 (10 to 21). On the cases where \toolname did not abstain, \toolname achieved higher sensitivity than the base MAS across all three codes (K74.0: 0.857 vs.\ 0.667; K74.6: 0.500 vs.\ 0.421; K76.0: 0.615 vs.\ 0.588), specificity comparable to the base MAS on K74.0 and K74.6 and slightly lower on K76.0 (0.375 vs.\ 0.435), and higher pooled accuracy (0.712 vs.\ 0.674). 

\begin{table}[tb]
\tbl{Per-class and overall comparison of MAS (no ethics) versus MAS $+$ ETHOS (3-class mode). N=50; Accuracy denotes decided accuracy (accuracy over non-abstained cases).\label{tab:mas-vs-ethos}}
{\fontsize{9}{10}\selectfont
\setlength{\tabcolsep}{6pt}
\renewcommand{\arraystretch}{1.15}
\begin{tabular}{@{}l cc cc cc cc@{}}
\toprule
 & \multicolumn{2}{c}{K74.0} & \multicolumn{2}{c}{K74.6} & \multicolumn{2}{c}{K76.0} & \multicolumn{2}{c}{Overall} \\
\cmidrule(lr){2-3}\cmidrule(lr){4-5}\cmidrule(lr){6-7}\cmidrule(lr){8-9}
Metric & Hep. MAS & $+$ETHOS & Hep. MAS & $+$ETHOS & Hep. MAS & $+$ETHOS & Hep. MAS & $+$ETHOS \\
\midrule
Abstain     & 17 (34\%)    & 21 (42\%)    & 13 (26\%)    & 17 (34\%)    & 10 (20\%)    & 21 (42\%)    & 40 (27\%)    & 59 (39\%)    \\
\midrule
Sensitivity & 0.667 & \textbf{0.857} & 0.421 & \textbf{0.500} & 0.588 & \textbf{0.615} & 0.559 & \textbf{0.657} \\
Specificity & 0.905 & \textbf{0.909} & 1.000 & \textbf{1.000} & \textbf{0.435} & 0.375 & \textbf{0.780} & 0.761 \\
F1          & 0.727 & \textbf{0.800} & 0.593 & \textbf{0.667} & 0.500 & \textbf{0.516} & 0.607 & \textbf{0.661} \\
Accuracy    & 0.818 & \textbf{0.897} & 0.703 & \textbf{0.758} & \textbf{0.500} & 0.483 & 0.674 & \textbf{0.712} \\
\bottomrule
\end{tabular}}
\end{table}

%% file: discussion.tex
\section{Discussion}
\label{sec:discussion}



\looseness=-1
In this paper, we proposed \toolname, a modular meta-agent ethics framework for clinical multi-agent systems (MAS) whose design is informed by clinical AI governance frameworks and the input of clinical stakeholders. We applied \toolname to a hepatology MAS that combines EHR laboratory values and CT imaging with retrieved clinical practice guidelines to screen for liver disease. Governance operated end to end, producing findings at agent tool call boundaries, during cross-agent evidence compilation, and at final adjudication. Across 150 decisions, \toolname improved sensitivity on all three ICD-10 diagnosis tasks and decided accuracy on two, while raising the share of decisions returned as indeterminate from 26.7\% to 39.3\%.

The governed system performs better on the cases it answers because it abstains on hard or ambiguous cases with insufficient evidence to support a diagnosis, where the ungoverned system still reports an answer. The accuracy gain is partly a selection effect rather than improved reasoning. Therefore, the relevant question is whether abstention is targeted at the right cases, not whether diagnostic accuracy improves. Our stratified analysis indicates that it is. When all required modalities were present and correctly matched, \toolname left the abstention rate unchanged for every code. When a required modality was missing or incorrect, abstention rose from 40\% to 62\%, driven primarily by the Imaging Appropriateness and Contextual checks. \toolname therefore withholds answers where it was designed to, and the resulting loss in coverage is a property of the design rather than a defect.

Building \toolname surfaced a design question we did not anticipate, namely where to draw the boundary between ethical requirements that hold across clinical applications and the operational checks that enforce them within a particular one. Our initial answer placed too much in the shared layer. An attempt to govern a second MAS that screens for cognitive impairment from patient speech found that no response could pass adjudication, because criteria specific to imaging had been written into the ethics critic rather than supplied as configuration. The intended invariant is that the critic adjudicates against the findings a system produces, not against a fixed list of the issues a particular application can raise. Recovering that invariant requires separating the critic's adjudication procedure, which is general, from the criteria it adjudicates against, which are not.

Several limitations bound these conclusions and present opportunities for future work. We govern a single MAS on a small, manually selected cohort, so the reuse we demonstrate is across the agents and tools of one system rather than across clinical applications. The concerns \toolname enforces are a subset of those our stakeholders raised, selected because they could be expressed as verifiable conditions. Finally, while diagnostic outputs were evaluated against ground-truth labels, the findings and adjudications \toolname produced were not reviewed by clinicians, so their clinical validity and their effect on clinician trust remain untested. Hence for future work we plan to govern additional MAS that differ in modality and task, operationalize a broader set of stakeholder concerns, and submit \toolname's findings for clinician adjudication.